\documentclass[aps,pra,reprint,superscriptaddress,longbibliography,floatfix]{revtex4-2}
\usepackage{amsmath,amssymb,bm,mathtools}
\usepackage{graphicx,booktabs}
\usepackage[nopatch=footnote]{microtype}
\usepackage[colorlinks=true,linkcolor=blue,citecolor=blue,urlcolor=blue,breaklinks=true]{hyperref}
\newcommand{\Tr}{\operatorname{Tr}}
\newcommand{\Rsec}{R_{\mathrm{sec}}}
\newcommand{\Jperp}{J_{\perp}}
\newcommand{\cD}{\mathcal D}
\hypersetup{pdftitle={Hilbert-space fragmentation selects channels of strong-to-weak spontaneous symmetry breaking},pdfauthor={Mingdi Xu, Kaixiang Lu, Daoguang Li, and Lei Pan}}
\begin{document}
\raggedbottom
\title{Hilbert-space fragmentation selects channels of strong-to-weak spontaneous symmetry breaking}
\author{Mingdi Xu}
\author{Kaixiang Lu}
\author{Daoguang Li}
\author{Lei Pan}
\email{panlei@nankai.edu.cn}
\affiliation{School of Physics, Nankai University, Tianjin 300071, China}
\begin{abstract}
Strong-to-weak spontaneous symmetry breaking (SWSSB) can occur in a maximally mixed charge sector even when ordinary charged correlations vanish. We study this order when a charge sector splits into dynamically disconnected fragments. In a one-dimensional $t$-$J_z$ chain with species-resolved dephasing, each fragment is labeled by a conserved spin word. A charged transfer between separated finite intervals preserves that word only if the intervening spins satisfy a periodicity constraint. In the full-state two-point sense, we use this constraint to prove the absence of long-range fidelity and R\'enyi-1 order in typical fixed-composition fragments, uniformly over all norm-bounded charged operators in any prescribed finite interval. A fragment with primitive spin-word period $k$ supports a $k$-particle channel with exact long-distance value $\nu^k(1-\nu)^k/k^2$, and no charged operator on fewer than $k$ sites can carry this order. A transverse exchange connects the fragments while preserving the global charges and restores the full-sector steady state at finite size. The conserved spin word therefore selects the local channels that can support mixed-state order.
\end{abstract}
\maketitle

\section{Introduction}

Strong-to-weak spontaneous symmetry breaking (SWSSB) distinguishes mixed states whose ordinary charged correlations vanish but whose nonlinear correlations remain long ranged~\cite{Sala2024Purification,Lessa2025SWSSB,Weinstein2025Renyi1,Gu2025OpenSSB}; Ref.~\cite{Wang2026Review} provides a recent review and broad bibliography. A state with definite Abelian charges has strong symmetry, whereas invariance under conjugation requires only weak symmetry. We use fidelity and R\'enyi-1 correlators to diagnose this order. The broader development grew out of symmetry-breaking descriptions of ergodic/delocalized dynamics and decoherence-induced mixed-state criticality~\cite{Garratt2021Delocalization,Lee2023Decoherence}, and now includes superoperator and hydrodynamic formulations~\cite{Ogunnaike2023Hydrodynamics,Moudgalya2024Superoperators,Huang2024Hydrodynamics}, as well as mixed-state phase frameworks based on separability, average symmetry, and finite Markov length~\cite{ChenGrover2024Separability,Ma2025AverageSymmetry,SangHsieh2025Markov}. SWSSB has also been studied in stochastic dephasing circuits~\cite{Kuno2024Dephasing} and decohered Ising and one-dimensional critical states~\cite{Orito2025MultipleDecoherence,Guo2025DecoheredCritical}. The connection between SWSSB and entanglement transitions has been studied in solvable complex Brownian Sachdev--Ye--Kitaev models~\cite{Chen2025EntanglementTransitions}. Maximally mixed states of complete charge sectors provide simple ordered examples~\cite{Ziereis2025SteadySWSSB}.

Hilbert-space fragmentation splits a symmetry sector into dynamically disconnected subspaces beyond those distinguished by global charges~\cite{Sala2020HSF,Khemani2020Shattering,Moudgalya2022HSF}. Related constrained dynamics appears in fractonic random circuits~\cite{Pai2019FractonicCircuits}, strictly confined systems~\cite{Yang2020StrictConfinement}, and strongly interacting models with fragmented Fock-space connectivity~\cite{DeTomasi2019FockFragmentation}. In open systems, fragments can preserve initial-state memory and support distinct stationary states~\cite{Li2023OpenHSF,Vuina2025Dephasing,Moudgalya2024Superoperators}; driven-dephasing Rydberg arrays provide another setting~\cite{Yan2025RydbergHSF}. Constraints also enter monitored-circuit studies of SWSSB with charge and dipole conservation~\cite{Zerba2025DipoleSWSSB}. Fixed-charge projectors with sparse random supports can also lack nonlinear order~\cite{Lee2026ChargeScrambling}. We ask how the nonlinear order of individual fragments differs when both their global charges and their dimensions are identical. An exponentially large support need not retain the ordered channels of the complete sector.

We study a one-dimensional $t$-$J_z$ chain with species-resolved dephasing. At zero transverse exchange, removing holes from an occupation configuration leaves a conserved spin word~\cite{Rakovszky2020Statistical}. This ordered-spin description is related to squeezed-space formulations of strongly correlated particles~\cite{OgataShiba1990,Kruis2004SqueezedSpace}. Spin-incoherent treatments use spin-sequence constraints to calculate ordinary one-body correlations~\cite{Fiete2004Anomalous,CheianovZvonarev2004,Matveev2007Bosonization}, pair correlations~\cite{Tilahun2008Pair}, and correlators of strongly interacting spinor gases~\cite{Patu2019TwoComponentCorrelators,Patu2024Spinor}. Here the object is instead a nonlinear correlator of a fixed-word projector. Word preservation turns elementary transfers into a configuration count and constrains every charged operator in a fixed finite interval.

The same constraint gives two results. Typical fixed-composition words have no long-range fidelity or R\'enyi-1 order in any such local channel. A primitive period-$k$ word instead supports an exact $k$-particle plateau, and no smaller local support can carry charged order. Perturbations can merge fragments in related $t$-$J_z$ settings~\cite{Lisiecki2025Tunable}. For the uniform transverse exchange and species-resolved dephasing considered here, our finite-size convergence proof establishes the full-sector endpoint without changing the global charges. This control separates the effect of the conserved word from that of global symmetry.

\begin{figure}[!t]
\centering
\includegraphics[width=\columnwidth]{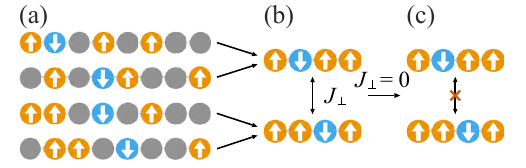}
\caption{Spin-word fragmentation on an open chain. (a) Orange and blue disks denote up and down spins; gray disks denote holes. Deleting the holes maps the four configurations onto two words. (b) Transverse spin exchange connects words with the same global charges. The word-space arrow denotes connectivity through hopping and transverse exchange; it need not be a direct transition between the displayed real-space configurations. (c) At $J_\perp=0$, hopping preserves each word. The red cross marks the forbidden transverse-exchange link. The illustrated sector has $L=8$, $N_\uparrow=3$, and $N_\downarrow=1$; it is separate from the balanced sector used in Fig.~\ref{fig:dynamics}.}
\label{fig:fragmentation}
\end{figure}

\section{Model, stationary states, and nonlinear diagnostics}
\label{sec:model}

\subsection{Spin-word fragmentation and stationary projectors}

We take the one-dimensional $t$--$J_z$ chain~\cite{BatistaOrtiz2000} and include transverse spin exchange as a control. Consider an open chain with local states $\{0,\uparrow,\downarrow\}$ and no double occupancy. Its Hamiltonian is
\begin{align}
H={}&-t_{\mathrm h}\sum_{i=1}^{L-1}\sum_\sigma
 \left(\widetilde c_{i\sigma}^{\dagger}\widetilde c_{i+1,\sigma}
 +\mathrm{H.c.}\right)
 +J_z\sum_{i=1}^{L-1}S_i^zS_{i+1}^z \nonumber\\
&+\Jperp\sum_{i=1}^{L-1}
 \left(S_i^+S_{i+1}^-+S_i^-S_{i+1}^+\right),
\label{eq:hamiltonian}
\end{align}
where the tilde enforces the occupancy constraint and
$S_i^z=(n_{i\uparrow}-n_{i\downarrow})/2$. We set $\hbar=t_{\mathrm h}=1$.
Species-resolved dephasing gives the Gorini--Kossakowski--Sudarshan--Lindblad master equation~\cite{Gorini1976,Lindblad1976}
\begin{align}
\dot\rho&=-i[H,\rho]+\gamma\sum_{i,\sigma}\cD[n_{i\sigma}]\rho,\nonumber\\
\cD[A]\rho&=A\rho A^\dagger-\tfrac12\{A^\dagger A,\rho\},
\label{eq:lindblad}
\end{align}
with $\gamma>0$. Both the Hamiltonian and every jump operator conserve
$N_\sigma=\sum_i n_{i\sigma}$. The strong
$U(1)_\uparrow\times U(1)_\downarrow$ symmetry~\cite{BucaProsen2012,AlbertJiang2014} is unchanged as $J_\perp$ changes.

For an occupation-basis configuration $\alpha$, define $\mathcal W(\alpha)$ by reading the occupied sites from left to right and deleting every hole. We call $\mathcal W(\alpha)=w=(w_1,\ldots,w_N)$ its spin word, where $N=N_\uparrow+N_\downarrow$ and $U,D$ abbreviate $\uparrow,\downarrow$. In Fig.~\ref{fig:fragmentation}(a), the first two configurations give $w=UDUU$, while the last two give $w=UUDU$. A hopping event exchanges a particle with a neighboring hole. It changes the occupied sites but cannot make two particles pass each other, so it leaves $w$ unchanged. The $J_z$ interaction and the dephasing jumps are diagonal in this basis and preserve $w$ as well. Thus at $J_\perp=0$ the fixed-charge Hilbert space splits into invariant word fragments. For each $w$, the $N$ ordered particles can occupy any $N$ sites, giving
\begin{equation}
d_w=\binom LN,\qquad
\mathcal N_w=\binom N{N_\uparrow}.
\label{eq:dimensions}
\end{equation}
Both are exponential in $L$ at fixed filling $\nu=N/L\in(0,1)$ and composition $p=N_\uparrow/N\in(0,1)$. All words of that composition have the same dimension. In thermodynamic limits, $N$ and $N_\uparrow$ depend on $L$, with $N/L\to\nu$ and $N_\uparrow/N\to p$.

Within a word, hopping connects any particle placement to the left-packed configuration and hence to every other placement. The hole pattern supplies no additional fragment label. Nonzero $J_\perp$ permits adjacent spin transpositions in a packed configuration, which generate every word of the same composition. The arrow in Fig.~\ref{fig:fragmentation}(b) denotes such connectivity through hopping and exchange, not necessarily a direct transition between the displayed configurations. At $J_\perp=0$ this interword path is absent [panel (c)].

Species-resolved dephasing removes coherences between occupation configurations. At finite size, stationary weights are constant on each connected component: the components are the individual word fragments at $J_\perp=0$ and the complete fixed-charge sector when $J_\perp\ne0$. For any initial state in that charge sector, the long-time state is
\begin{equation}
\rho(t)\xrightarrow[t\to\infty]{}
\begin{cases}
\displaystyle\sum_w p_w\rho_w,&J_\perp=0,\\[3pt]
\rho_{\rm sec},&J_\perp\ne0,
\end{cases}
\label{eq:steady_states}
\end{equation}
where $\rho_0=\rho(0)$, $p_w=\Tr(\Pi_w\rho_0)$,
$\rho_w=\Pi_w/d_w$, and
$\rho_{\rm sec}=\Pi_{N_\uparrow,N_\downarrow}/(d_w\mathcal N_w)$.
The projectors select the word fragment and the complete fixed-charge sector, respectively. Appendix~\ref{app:fixedpoints} proves convergence as well as the stationary-state classification. Equation~\eqref{eq:steady_states} takes the long-time limit at fixed $L$; it does not impose a size-independent relaxation time. Below we study individual $\rho_w$. Their nonlinear correlators cannot be averaged to obtain those of a general mixture $\sum_w p_w\rho_w$.

\subsection{Nonlinear correlators as configuration counts}

Let $O_x$ have definite nonzero species charge $\bm q=(q_\uparrow,q_\downarrow)$, defined by $[N_\sigma,O_x]=q_\sigma O_x$. Its signed particle-number change is $\delta=q_\uparrow+q_\downarrow$. The neutral transfer between separated regions is $X_{x,y}=O_y^\dagger O_x$. We use
\begin{align}
F_O&=\Tr\sqrt{\sqrt\rho X_{x,y}\rho X_{x,y}^\dagger\sqrt\rho},\nonumber\\
R_{1,O}&=\Tr\!\left[X_{x,y}\sqrt\rho X_{x,y}^\dagger\sqrt\rho\right]
\label{eq:diagnostics}
\end{align}
as the fidelity and R\'enyi-1 correlators, with $F_O=F_{X_{x,y}}$ and $R_{1,O}=R_{1,X_{x,y}}$ in the transfer notation of Appendix~\ref{app:projector}. Fidelity is unsquared, and $X\rho X^\dagger$ is not separately normalized. In the projector states considered here, ordinary correlators of charged operators on disjoint regions vanish because $X_{x,y}$ has no diagonal matrix element. Long-range nonlinear order can nevertheless survive.

For $\rho=\Pi/d$, set $A=\Pi X_{x,y}\Pi$. Then
\begin{equation}
F_O=\frac{\|A\|_1}{d},\qquad
R_{1,O}=\frac{\|A\|_2^2}{d},\qquad F_O^2\le R_{1,O},
\label{eq:projector_count}
\end{equation}
where $\|\cdot\|_1$ and $\|\cdot\|_2$ are the trace and Hilbert--Schmidt norms \cite{Weinstein2025Renyi1,Liu2024Wightman}. Suppose $\Pi$ selects a configuration set $\Omega$, and $X$ maps distinct nonzero inputs to distinct configuration outputs with unit-modulus amplitudes. Then $A$ has only zero or unit singular values, giving
\begin{align}
F_O=R_{1,O}&=\frac{N_X}{d},\qquad d=|\Omega|,\nonumber\\
N_X&=\#\bigl\{\alpha\in\Omega:\ X|\alpha\rangle\ne0,\nonumber\\[-2pt]
&\hspace{28mm}\Pi X|\alpha\rangle=X|\alpha\rangle\bigr\}.
\label{eq:success_count}
\end{align}
The R\'enyi-1 expression is the successful-transition count for diagonal projectors used in Ref.~\cite{Lee2026ChargeScrambling}. Fermionic signs do not change this count. Success means that the output remains in the original support; it need not equal the input. The denominator includes configurations annihilated by $X$. General coherent operators need the norm relations in Eq.~\eqref{eq:projector_count}, rather than a configuration fraction (Appendix~\ref{app:projector}).

In the complete charge sector, a one-particle up-spin transfer
$X_{i,r}^{(1)}=\widetilde c_{i+r,\uparrow}^\dagger\widetilde c_{i,\uparrow}$ succeeds whenever its source contains an up spin and its target is empty. Here $(i,r)$ denotes source and displacement, corresponding to $(x,y)=(i,i+r)$. At any nonzero separation,
\begin{equation}
F_{\rm sec}^{(1)}=R_{1,\rm sec}^{(1)}
=\Rsec(L)=\frac{N_\uparrow(L-N)}{L(L-1)}
\longrightarrow \nu p(1-\nu).
\label{eq:sector_value}
\end{equation}
The nonzero limit and vanishing ordinary correlator establish sector SWSSB. At half filling and spin balance the limit is $1/8$.

The complete sector imposes only charge conservation; a single fragment also requires preservation of the spin word.

\section{Absence of two-point order in typical fragments}
\label{sec:typical}

\begin{figure}[!t]
\centering
\includegraphics[width=\columnwidth]{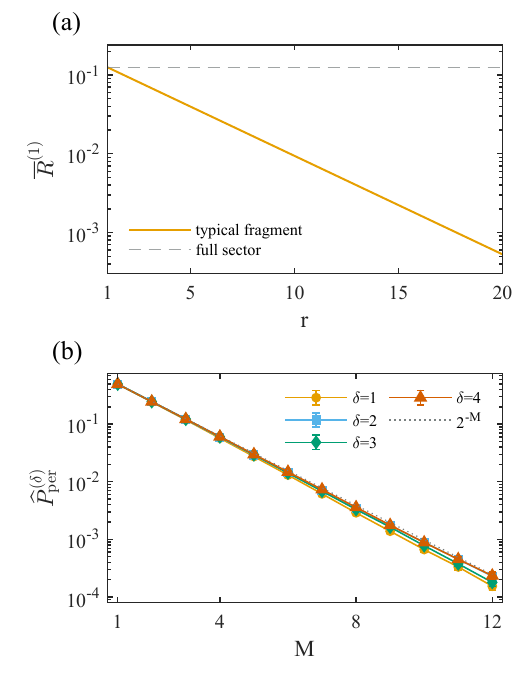}
\caption{Suppression of local nonlinear order in typical fragments. (a) Thermodynamic analytic single-particle result $\tfrac18(3/4)^{r-1}$ and complete-sector reference $1/8$ at $\nu=p=1/2$. (b) Mean fraction $\widehat P_{\rm per}^{(\delta)}$ of rank windows satisfying $M$ consecutive equalities $w_j=w_{j+\delta}$, from $5000$ uniformly sampled balanced words of length $N=128$, with $M=1,\ldots,12$ and $\delta=1,2,3,4$. Bars give the standard error across words, and solid lines connect neighboring integer values of $M$ as guides to the eye. The dotted reference $2^{-M}$ is the balanced $N\to\infty$ limit; the exact finite-$N$ fixed-composition mean is given by Eq.~\eqref{eq:app-finiteNperiodicity}. This statistic tests a necessary word constraint and differs from the correlator in (a). The exclusion of all fixed finite local charged channels follows from Eq.~\eqref{eq:typical_bound}.}
\label{fig:typical}
\end{figure}

A successful transfer must preserve the complete spin word. For a one-particle up-spin transfer, every particle passed must be an up spin, so the count is controlled by same-spin runs. An overbar denotes the average of the correlator over allowed real-space source positions; $\mathbb E_w$ and $\Pr_w$ refer instead to the word ensemble. Appendix~\ref{app:counting} derives the exact source average for any finite word. For typical words of composition $p$,
\begin{equation}
\frac{\overline R_w^{(1)}(r;L)}{\Rsec(L)}
\xrightarrow[L\to\infty]{\rm prob.}
(1-\nu+\nu p)^{r-1}.
\label{eq:onebody_typical}
\end{equation}
Each intervening site is either empty or contains a compatible up spin. At half filling and spin balance this gives $\overline R_{\rm typ}^{(1)}(r)=\tfrac18(3/4)^{r-1}$ [Fig.~\ref{fig:typical}(a)]. The one-particle channel is short ranged, but a statement about every local channel requires a more general count.

Fix an enclosing interval length $K$, and place $O_x$ and its translate $O_{x+r}$ in the ordered intervals $I_x=[x,x+K-1]$ and $I_{x+r}$, with $r\ge K$. The untouched gap has $g=r-K$ sites. We require $\|O\|\le1$ in operator norm and definite nonzero species charge. If $\delta=0$, word conservation forces the projected transfer to vanish: the particle rank at a cut through the gap is unchanged, but the species content of the left prefix has changed. The right operator cannot repair that prefix. This argument uses left-right separation of the enclosing intervals.

For $\delta\ne0$, the transfer shifts the ranks of all untouched gap particles by $\delta$. A necessary condition for remaining in the same fragment is
\begin{equation}
w_j=w_{j+\delta}
\quad\text{for every gap particle.}
\label{eq:periodicity}
\end{equation}
Here $\delta$ is a signed particle-number change, not a spatial separation. Endpoint matching is also required; Appendix~\ref{app:generallocal} gives the full rank-shift geometry. In a locally Bernoulli word, $M$ such equalities form disjoint chains. With $\theta=\max(p,1-p)$, the probability that all chains are monochromatic is at most $\theta^M$. The limiting gap occupation is $M_g\sim\operatorname{Bin}(g,\nu)$, so
\begin{equation}
\mathbb E\!\left[\theta^{M_g}\right]
=(1-\nu+\nu\theta)^{r-K}
\equiv\lambda^{r-K},\qquad \lambda<1.
\label{eq:lambda}
\end{equation}
The exponential decay combines the probability of matching spins in particle-rank space with the particle-number distribution in the real-space gap.

Let $\mathcal O_K$ contain all norm-bounded charged operators supported within a reference interval of length $K$, with the neutral component removed when several charge sectors are present.  For a fixed word and pair of intervals, let $e_w$ be the fraction of input configurations for which at least one signed displacement $0<|\delta|\le K$ satisfies the rank-matching condition~\eqref{eq:periodicity} across the entire gap.  Every projected charged transfer has nonzero columns only on this common eligible set, while its operator norm is at most unity.  Consequently both its fidelity and R\'enyi-1 correlators are bounded by $e_w$, uniformly over $O\in\mathcal O_K$.  A union bound over the at most $2K$ signed displacements and Eq.~\eqref{eq:lambda} then give, for every $\varepsilon>0$,
\begin{align}
\limsup_{L\to\infty}\Pr_w\!\left[
\sup_{O\in\mathcal O_K}\overline F_{O,L}(r;w)>\varepsilon
\right]
&\le\frac{2K}{\varepsilon}\lambda^{r-K},\nonumber\\
\limsup_{L\to\infty}\Pr_w\!\left[
\sup_{O\in\mathcal O_K}\overline R_{1,O,L}(r;w)>\varepsilon
\right]
&\le\frac{2K}{\varepsilon}\lambda^{r-K}.
\label{eq:typical_bound}
\end{align}
Here the words are sampled uniformly at fixed composition.  The same exponential suppression holds at prescribed macroscopic bulk sources chosen independently of the word.  At fixed $K$, the right-hand sides vanish as $r\to\infty$, proving the absence of conventional full-state two-point SWSSB in typical fragments for every prescribed finite local support.  Appendix~\ref{app:generallocal} gives the common-input rank proof and the source-averaged almost-sure statement.

Figure~\ref{fig:typical}(b) tests the word statistics behind the bound. Their suppression with $M$ becomes spatial decay after averaging over the gap occupancy. The fixed-$M$ scan in Appendix~\ref{app:numerics}, evaluated at $N=16,32,\ldots,256$, instead approaches the nonzero value $2^{-M}$; Appendix~\ref{app:numerics} also gives the exact finite-$N$ fixed-composition reference. Equation~\eqref{eq:typical_bound} keeps $K$ fixed, takes $L\to\infty$ first, and then takes $r\to\infty$. Source selection cannot adapt to rare windows found in a word. Here ``local'' refers to the operator support; the fidelity is evaluated on the full fragment state. This differs from reduced-state local-fidelity definitions~\cite{Divi2026Local} and should be distinguished from other local formulations of strong symmetry and its breaking~\cite{Liu2026LocalStrongSymmetry}. Periodic words can satisfy the same matching condition in selected composite channels without a rare spin fluctuation.

\section{Composite SWSSB and minimum operator support}
\label{sec:periodic}

\begin{figure}[!t]
\centering
\includegraphics[width=\columnwidth]{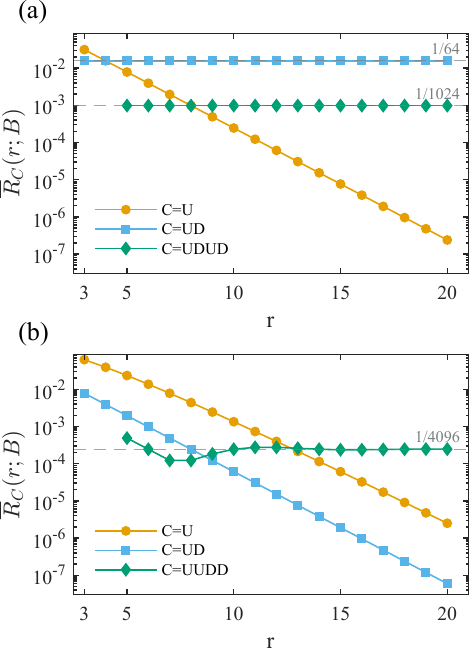}
\caption{Primitive spin-word periods select composite SWSSB channels. Markers show exact thermodynamic source-averaged configuration counts at $\nu=p=1/2$. (a) In $B=UD$, the $C=U$ channel decays, while $C=UD$ and $UDUD$ have plateaus $1/64$ and $1/1024$. The minimum ordered support is $K_{\min}=2$. (b) In $B=UUDD$, the same $U$ and $UD$ channels decay, whereas $C=UUDD$ approaches $1/4096$ and $K_{\min}=4$. Solid lines connect neighboring integer separations only as guides to the eye, and dashed lines mark the analytic plateaus. The oscillation in (b) follows from exact gap counting; these data have no sampling uncertainty. The background period is $k=|B|$ and the transferred motif length is $\ell=|C|$. The markers are shown for integer separations $r=\max(3,\ell+1),\ldots,20$; contact values appear in Appendix~\ref{app:periodic}.}
\label{fig:periodic}
\end{figure}

Let the spin word repeat a primitive block $B=(\sigma_1,\ldots,\sigma_k)$ of length $k$, so $w=B B\cdots B$. A transferred motif $C$ of length $\ell$ need not be a primitive block. In the alternating background $B=UD$, for example, a single up-spin transfer has $\overline R_U(r;UD)=2^{-r-2}$ at half filling, while transferring $C=UD$ gives $1/64$ for every $r\ge3$ [Fig.~\ref{fig:periodic}(a)]. Here $\overline R_C(r;B)$ denotes the common source-averaged fidelity and R\'enyi-1 count after $L\to\infty$ at fixed $r$.

For a primitive-block transfer, define
\begin{equation}
O_i^{(B)}=\widetilde c_{i+k-1,\sigma_k}\cdots\widetilde c_{i,\sigma_1},
\quad X_{i,r}^{(B)}=O_{i+r}^{(B)\dagger}O_i^{(B)}.
\label{eq:periodic_block}
\end{equation}
For $r\ge k$, success requires an occupied source with motif $B$, an empty target, and a gap particle count divisible by $k$. The source phase occurs with probability $1/k$, giving
\begin{align}
\overline R_B(r;B)
&=\frac{\nu^k(1-\nu)^k}{k}
\Pr[M_{r-k}=0\pmod{k}],\nonumber\\
\lim_{r\to\infty}\overline R_B(r;B)
&=\frac{\nu^k(1-\nu)^k}{k^2}>0,
\label{eq:periodic_order}
\end{align}
where $M_{r-k}\sim\operatorname{Bin}(r-k,\nu)$. The second factor $1/k$ in the plateau comes from the asymptotically uniform residues of the gap particle count, without assuming independent endpoint phases. Since the ordinary correlator vanishes, this is a composite SWSSB channel. Appendix~\ref{app:periodic} gives the finite-distance roots-of-unity expansion and the distinct contact value at $r=k$.

The primitive period also sets the smallest ordered support:
\begin{equation}
\delta\in k\mathbb Z\setminus\{0\}\ \text{is necessary},
\qquad K_{\min}=k.
\label{eq:minimum_support}
\end{equation}
If $\delta$ is not a multiple of $k$, a gap containing at least $k$ particles encounters an incompatible spin in Eq.~\eqref{eq:periodicity}. The probability of avoiding that mismatch vanishes with distance. Zero-$\delta$ transfers with nonzero species charge vanish exactly. Fewer than $k$ support sites cannot produce an allowed nonzero particle-number change, while Eq.~\eqref{eq:periodic_block} attains the lower bound. This argument also bounds sums of nonzero-charge components (Appendix~\ref{app:periodic}). Divisibility alone is not sufficient for an arbitrary motif; the endpoint occupations and word phase must match.

Figure~\ref{fig:periodic} compares the same $UD$ pair in two balanced backgrounds. It is ordered for $B=UD$ but decays as $2^{-r-4}$ for $B=UUDD$. The latter supports a four-particle plateau with damped oscillations from the exact roots-of-unity sum. Changing the primitive period therefore changes the minimum ordered support as well as the plateau. The $UDUD$ motif in panel (a) still belongs to a period-two background. These results classify the primitive-period family; they do not require every word with nonlinear order to be globally periodic.

The periodic constraint also selects the species-charge direction.  If the primitive block has content $\boldsymbol b=(n_\uparrow(B),n_\downarrow(B))$, then a word-preserving ordered process with displacement $\delta=mk$ must carry species charge $\boldsymbol q=m\boldsymbol b$.  Thus the ordered channels of $UD$ and $UUDD$ backgrounds lie along the total-particle directions generated by $(1,1)$ and $(2,2)$, respectively.  By contrast, a pure relative-spin transfer such as $O_i=S_i^+$ has $\boldsymbol q=(1,-1)$ and $\delta=0$, and its projected two-point transfer vanishes in every fixed-word fragment.  In the complete fixed-charge projector the same channel has $F_{S^+}=R_{1,S^+}=N_\uparrow N_\downarrow/[L(L-1)]\to\nu^2p(1-p)$, equal to $1/16$ at half filling and spin balance.  Appendix~\ref{app:periodic} gives the prefix-counting derivation.

\begin{figure}[!t]
\centering
\includegraphics[width=\columnwidth]{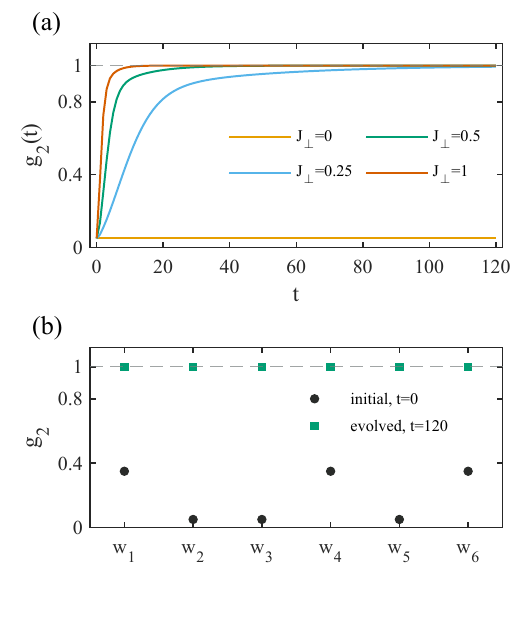}
\caption{Finite-size restoration of sector correlations by transverse exchange. Parameters are $L=8$, $N_\uparrow=N_\downarrow=2$, $t_{\mathrm h}=1$, $J_z=0.7$, and $\gamma=1$, with source-target separation $r=4$; $g_2$ uses $\Rsec(8)=1/7$. (a) The curves use $J_\perp=0,0.25,0.5,1$. Starting from $\rho_{UDUD}$, $g_2=1/20$ stays constant at $J_\perp=0$, while nonzero exchange drives it toward $1$. Each solid curve joins 121 density-matrix evaluations at $t=0,1,\ldots,120$. (b) Black dots show $g_2(0)$ and green squares show $g_2(120)$ at $J_\perp=1$, for $(w_1,\ldots,w_6)=(UUDD,UDUD,UDDU,DUUD,DUDU,DDUU)$. Their initial values depend on the fragment, whereas the evolved values approach the same sector reference. These finite-size data illustrate Eq.~\eqref{eq:steady_states}; Eq.~\eqref{eq:sector_value} gives the thermodynamic sector value.}
\label{fig:dynamics}
\end{figure}

\section{Symmetry-preserving restoration of sector order}
\label{sec:restoration}

Nonzero $J_\perp$ connects words while conserving both species charges. Equation~\eqref{eq:steady_states} then selects the full-sector state at fixed $L$. Figure~\ref{fig:dynamics} illustrates the restoration of its single-particle channel for $L=8$, $N_\uparrow=N_\downarrow=2$, $J_z=0.7$, and $\gamma=1$.

During evolution, we use the R\'enyi-2 diagnostic and its sector-normalized source average,
\begin{align}
R_{2,X}(t)&=\frac{\Tr[X\rho(t)X^\dagger\rho(t)]}{\Tr[\rho(t)^2]},\nonumber\\
g_2(t)&=\frac1{\Rsec(L)(L-r)}\sum_{i=1}^{L-r}R_{2,X_{i,r}^{(1)}}(t),
\quad r=L/2.
\label{eq:g2}
\end{align}
For these configuration transfers, $F=R_1=R_2$ at uniform-projector endpoints. Away from them, $R_2$ is a transient diagnostic and need not equal the other two quantities~\cite{Lessa2025SWSSB,Chen2025EntanglementTransitions}. We evaluate it directly from the density matrix. Multi-copy protocols access polynomial state functionals~\cite{Ekert2002NonlinearFunctionals,Brun2004PolynomialFunctions}, and randomized-measurement schemes have been proposed for SWSSB R\'enyi-2 diagnostics~\cite{Sun2025Randomized}.

The alternating projector is stationary at zero exchange and gives $g_2=1/20$. With exchange, it approaches $g_2=1$ [Fig.~\ref{fig:dynamics}(a)]. All six balanced words approach that same reference despite their different initial values [panel (b)]. This restores the full-sector single-particle value; the alternating fragment already has composite order. The ordinary-correlator check and numerical methods are in Appendix~\ref{app:numerics}. These finite-size evolutions do not determine how the mixing time or spectral gap scales with $L$.

\section{Discussion}

Fragments with identical charges and dimensions can support different channels of nonlinear order. Lee's random-support construction already gives fixed-charge projectors without long-range R\'enyi-1 order for fixed finite-support charge transfers~\cite{Lee2026ChargeScrambling}. Our support contains every particle placement compatible with one conserved spin word. These correlations between configurations determine which transfers survive: typical words obey a uniform finite-interval bound, while periodic words of the same composition and dimension support matched composite channels with minimum support equal to their primitive period. Appendix~\ref{app:projector} compares random subsets and word fragments at the same rank, showing that exponential support size by itself is not the distinction: the internal correlations imposed by the conserved word determine the surviving transfer graph. The word constraint also relates these results to spin-sequence selection rules for ordinary correlations~\cite{Tilahun2008Pair,Patu2024Spinor}; complexity-based limitations concern more general SWSSB diagnostics~\cite{Feng2025Hardness}.

Transverse exchange removes the word constraint while preserving the strong global symmetry. The static channel results and finite-size endpoints are exact, but connectivity does not determine the time required to reach that endpoint. Links between SWSSB, hydrodynamics, and Lindbladian low-energy structure~\cite{Huang2024Hydrodynamics,Ogunnaike2023Hydrodynamics,Moudgalya2024Superoperators}, and studies of nonlinear-order growth in other open systems~\cite{Shu2026Scaling,Hauser2026Hydrodynamics}, motivate a separate dynamical analysis. Fragmentation robustness~\cite{Han2026Robustness} and word-constrained relaxation~\cite{Balasubramanian2024GlassyWords} likewise distinguish connectivity from relaxation. Metastability~\cite{Macieszczak2016Metastability} and the contrast between algebraic and exponential decoherence~\cite{Cai2013AlgebraicDecoherence} offer useful frameworks, without fixing an exponent for this model. The dependence of the mixing time on $L$ and $J_\perp$, and the classification of more general nonperiodic words, remain open.

Cold-atom experiments have explored constrained dynamics in tilted Fermi--Hubbard chains~\cite{Scherg2021Nonergodicity,Kohlert2023Fragmentation} and hidden spin correlations through string correlators~\cite{Hilker2017StringCorrelations}. Mixed-state phases have been probed on a quantum computer~\cite{Zhang2026Probing}, while SWSSB has been observed in a dephased Fermi gas~\cite{Wang2026Experiment}. Preparing the two-species projectors and probing their periodic composite channels would require a dedicated protocol. Randomized-measurement methods access nonlinear moments~\cite{Elben2018Random,Elben2019StatisticalToolbox}, including experimental R\'enyi-entropy measurements~\cite{Brydges2019RandomizedRenyi}. Measuring the present charge-transfer correlators still requires treating the nonunitary operators and their sampling cost.

\begin{acknowledgments}
This work was supported by the National Natural Science Foundation of China under Grant Nos.~12304290 and 12505017, and by the Beijing National Laboratory for Condensed Matter Physics under Grant No.~\mbox{2025BNLCMPKF017}. L.P. also acknowledges support from the Fundamental Research Funds for the Central Universities (010-63263115).
\end{acknowledgments}

\label{maintext:end}

\appendix

\section{Projector states and configuration counts}
\label{app:projector}

Let $\Pi$ be an orthogonal projector of rank $d$ and set $\rho=\Pi/d$.
For an arbitrary transfer operator $X$, write $A=\Pi X\Pi$.
For $X_{x,y}=O_y^\dagger O_x$, the local-operator notation means
$F_O(x,y;\rho)=F_{X_{x,y}}(\rho)$ and
$R_{1,O}(x,y;\rho)=R_{1,X_{x,y}}(\rho)$.
Relative to the squared convention for mixed-state fidelity~\cite{Jozsa1994}, we use the root (unsquared) functional. For a nonunitary transfer, $X\rho X^\dagger$ need not be normalized, and we do not renormalize it.
The Schatten norms are $\|A\|_p=(\sum_j s_j^p)^{1/p}$, where $s_j$
are the singular values of $A$.  Substituting $\sqrt\rho=\Pi/\sqrt d$
into the two nonlinear diagnostics gives
\begin{align}
 F_X&=\frac1d\Tr\sqrt{AA^\dagger}=\frac{\|A\|_1}{d},\nonumber\\
 R_{1,X}&=\frac1d\Tr(AA^\dagger)=\frac{\|A\|_2^2}{d}.
 \label{eq:app-projector}
\end{align}
Since $\operatorname{rank}A\le d$, Cauchy--Schwarz implies
\begin{equation}
 F_X^2=\frac{(\sum_j s_j)^2}{d^2}
 \le\frac{\sum_j s_j^2}{d}=R_{1,X}.
 \label{eq:app-fidelitybound}
\end{equation}
This argument places no restriction on the matrix elements of $X$.
If $\|X\|\le1$, then $s_j\le1$ also gives $R_{1,X}\le F_X$.
The normalized R\'enyi-2 diagnostic obeys
\begin{equation}
 R_{2,X}=\frac{\Tr(X\rho X^\dagger\rho)}{\Tr\rho^2}
 =\frac{\|A\|_2^2}{d}=R_{1,X}
 \label{eq:app-r2projector}
\end{equation}
for every uniform projector state.  During a generic transient, the
three diagnostics must be evaluated separately.

Suppose $\Pi$ selects a subset $\Omega$ of the configuration basis
and that $X$ sends each applicable configuration to a distinct configuration
with unit-modulus amplitude.  The projected transfer $A$ is then a partial
isometry: its singular values are zero or one.
Let $N_X$ count the initial configurations $\alpha\in\Omega$ for which
$X|\alpha\rangle$ is nonzero and remains in $\operatorname{ran}\Pi$.
Then
\begin{equation}
 F_X=R_{1,X}=R_{2,X}=\frac{N_X}{|\Omega|}.
 \label{eq:app-successcount}
\end{equation}
Fermionic signs leave this count unchanged.  A general coherent local
operator need not be such a transfer; its matrix-unit expansion will be
bounded in Appendix~\ref{app:generallocal}.

For comparison, take the random-support ensemble used in
Ref.~\cite{Lee2026ChargeScrambling}: choose $\Omega$ uniformly among all
$d$-element subsets of a fixed-charge configuration sector of dimension
$D>1$.  For an operator $X$ with no diagonal matrix elements, each
ordered pair of distinct configurations belongs to $\Omega$ with
probability $d(d-1)/[D(D-1)]$.  Hence
\begin{equation}
 \mathbb E_\Omega R_{1,X}(\Pi_\Omega/d)
 =\frac{d-1}{D-1}\,R_{1,X}(\Pi_{\rm sec}/D).
 \label{eq:app-randomsupport}
\end{equation}
The suppression depends on the relative support size and can persist
even when $d$ grows exponentially.  At the rank of a word fragment,
$d=d_w$ and $D=d_w\mathcal N_w$, the prefactor is asymptotic to
$1/\mathcal N_w$ and is exponentially small at fixed filling and
composition.  A word fragment contains all placements of a prescribed
spin sequence, so its configurations are not a uniform random subset.
Their correlated transfer structure permits the periodic plateaus in
Appendix~\ref{app:periodic} at that same rank.

For disjoint source and target regions, a transfer of a nonzero species
charge changes at least one local occupation label.  It has no diagonal
configuration-basis matrix element.  Thus its ordinary correlator vanishes
in every diagonal fragment state, even when the nonlinear count is finite.
For the complete charge-sector projector, a single up-spin transfer succeeds
precisely when the source is up and the target is empty.  Sampling these two
sites without replacement gives
\begin{equation}
 \Rsec(L)
 =\frac{N_\uparrow}{L}\frac{L-N}{L-1}
 \longrightarrow\nu p(1-\nu).
 \label{eq:app-sectorone}
\end{equation}
Here $N=N_\uparrow+N_\downarrow$.  Thermodynamic limits use
$N=N(L)$ and $N_\uparrow=N_\uparrow(L)$, with $N/L\to\nu$ and
$N_\uparrow/N\to p$.  At $\nu=p=1/2$ the limit is $1/8$.

\section{Finite-size stationary states and convergence}
\label{app:fixedpoints}

We work in one fixed $(N_\uparrow,N_\downarrow)$ sector of an open chain,
with $t_{\mathrm h}\ne0$ and $\gamma>0$. Classical results describe irreducibility, stationary states, and their domains of attraction~\cite{Evans1977,Frigerio1978}; recent developments are discussed in Ref.~\cite{Yoshida2026SteadyStates}. The following model-specific commutant argument also proves finite-size convergence.  For the species-resolved bath,
the generator acting on any operator $Y$ is
\begin{equation}
 \mathcal L(Y)=-i[H,Y]
 -\frac\gamma2\sum_{i,\sigma}[n_{i\sigma},[n_{i\sigma},Y]].
 \label{eq:app-generator}
\end{equation}
Taking the real part of its Hilbert--Schmidt quadratic form gives
\begin{equation}
 \operatorname{Re}\Tr[Y^\dagger\mathcal L(Y)]
 =-\frac\gamma2\sum_{i,\sigma}\|[n_{i\sigma},Y]\|_2^2.
 \label{eq:app-dissipation}
\end{equation}
The Hamiltonian contribution has zero real part.  If $Y$ is stationary,
Eq.~\eqref{eq:app-dissipation} forces it to commute with every
$n_{i\sigma}$.  Their joint eigenvalues distinguish all allowed
configurations, so $Y=\sum_\alpha y_\alpha|\alpha\rangle\langle\alpha|$.
The remaining condition is
\begin{equation}
 [H,Y]_{\alpha\beta}=H_{\alpha\beta}(y_\beta-y_\alpha)=0.
 \label{eq:app-graphcondition}
\end{equation}
The coefficients of $Y$ are constant on each connected component of the graph whose edges
are nonzero off-diagonal Hamiltonian matrix elements.  Conversely, a
diagonal operator constant on every component is stationary. Graph-theoretic approaches also provide a complementary route to detecting Hilbert-space fragmentation~\cite{Rutkowski2026GraphFragmentation}; the components here follow directly from the displayed commutator.

At $J_\perp=0$, hopping only exchanges a particle with a neighboring hole
and preserves the ordered spin word.  Any configuration with word $w$ can
be packed against the left boundary by successive allowed hops; reversing
such a sequence reaches any other placement of the same word.  Components
are therefore exactly the word fragments, each with $d_w=\binom LN$
configurations.  For $J_\perp\ne0$, a packed configuration also permits
adjacent exchanges of unlike spins.  Such exchanges generate every binary
word with the given composition.  Packing, exchanging, and unpacking
connect any two configurations in the complete charge sector, whose
dimension is $d_{\mathrm{sec}}=\binom LN\binom N{N_\uparrow}$.
The diagonal Ising interaction leaves these graph connections unchanged.

General semigroup results distinguish stationary-state structure, invariant subspaces, and attraction to the asymptotic space~\cite{BaumgartnerNarnhofer2008,Spohn1977ApproachEquilibrium,Ticozzi2008MarkovianSubsystems}. To prove convergence at finite size, we exclude nondecaying modes outside the stationary subspace.
If $\mathcal L(Y)=i\omega Y$ for real $\omega$,
Eq.~\eqref{eq:app-dissipation} again makes $Y$ diagonal.
Then $-i[H,Y]$ is off diagonal, whereas $i\omega Y$ is diagonal.
Both must vanish, and a nonzero $Y$ requires $\omega=0$.
A finite-dimensional completely positive trace-preserving semigroup is
uniformly bounded, so a Jordan block at an eigenvalue on the imaginary
axis is also excluded: it would produce polynomial growth.
All remaining modes decay.  The conserved component weights fix
\begin{align}
 \lim_{t\to\infty}\rho(t)
 &=\sum_w\Tr(\Pi_w\rho_0)\frac{\Pi_w}{d_w},
 &&J_\perp=0,\nonumber\\
 \lim_{t\to\infty}\rho(t)
 &=\frac{\Pi_{N_\uparrow,N_\downarrow}}{d_{\mathrm{sec}}},
 &&J_\perp\ne0.
 \label{eq:app-endpoints}
\end{align}
The long-time map is an asymptotic projection whose stationary-component weights are determined by conserved functionals~\cite{Albert2016GeometryResponse}. These statements require finite $L$ and give no size-independent relaxation
time or gap.  For total-density dephasing, the local occupations do not
distinguish the spin labels, so the first step of this proof does not
apply; results for that bath are separate finite-size controls.

\section{Single-particle transfer for an arbitrary word}
\label{app:counting}

Consider $X_{i,r}^{(1)}=\widetilde c_{i+r,\uparrow}^\dagger
\widetilde c_{i,\uparrow}=X_{i,i+r}|_{O=\widetilde c_\uparrow}$
and average over $i=1,\ldots,L-r$.  An overbar denotes this real-space
source average, taken after evaluating each correlator.
Suppose the source has particle rank $j$ and the spatial interval strictly
between the endpoints contains $m$ particles.  The transfer changes the
subword $(w_j,w_{j+1},\ldots,w_{j+m})$ into
$(w_{j+1},\ldots,w_{j+m},w_j)$.  With $w_j=\uparrow$, equality of the
two words requires every spin in this subword to be up.  Define
\begin{align}
 A_m(w)&=\sum_{j=1}^{N-m}
 \mathbf1[w_j=\cdots=w_{j+m}=\uparrow]\nonumber\\
 &=\sum_s\max(a_s-m,0),
 \label{eq:app-runcount}
\end{align}
where $a_s$ are the up-run lengths.

For fixed $j,m,i$, the source and target occupations are fixed, while
the other particles can be placed in
\begin{equation}
 \binom{i-1}{j-1}\binom{r-1}{m}
 \binom{L-i-r}{N-j-m}
 \label{eq:app-fixedsourcecount}
\end{equation}
ways.  A binomial coefficient is zero outside its allowed range.
The convolution
\begin{equation}
 \sum_{i=1}^{L-r}\binom{i-1}{j-1}
             \binom{L-i-r}{N-j-m}
 =\binom{L-r}{N-m}
 \label{eq:app-runvandermonde}
\end{equation}
is independent of $j$.  Summing over allowed ranks and dividing by the
fragment dimension and the number of sources gives the exact result
\begin{equation}
 \overline R_w^{(1)}(r;L)=
 \frac{\displaystyle\sum_{m=0}^{\min(r-1,N-1)}
 A_m(w)\binom{r-1}{m}\binom{L-r}{N-m}}
 {\displaystyle(L-r)\binom LN}.
 \label{eq:app-runformula}
\end{equation}
It is the common source-averaged value of $F$, $R_1$, and $R_2$
for this configuration transfer.

\subsection{Thermodynamic run-survival criterion}
\label{app:runtransform}

For $N_\uparrow>0$, put $s_m^{(L)}=A_m/N_\uparrow$.  This is the
probability that an up spin selected uniformly from the word is followed
by at least $m$ further up spins.  Suppose $s_m^{(L)}\to s_m$ for every
fixed $m$.  The sequence $s_m$ decreases from $s_0=1$ to a limit
$s_\infty\ge0$.  Using the falling factorial $(x)_a=x(x-1)\cdots(x-a+1)$,
\begin{equation}
 \frac{\binom{L-r}{N-m}}{\binom LN}
 =\frac{(N)_m(L-N)_{r-m}}{(L)_r}
 \longrightarrow\nu^m(1-\nu)^{r-m}.
\end{equation}
At fixed $r$, division by Eq.~\eqref{eq:app-sectorone} then gives
\begin{align}
 \lim_{L\to\infty}\frac{\overline R_w^{(1)}(r;L)}{\Rsec(L)}
 &=\sum_{m=0}^{r-1}\binom{r-1}{m}
       \nu^m(1-\nu)^{r-1-m}s_m\nonumber\\
 &=\mathbb E[s_{M_r}],\qquad
 M_r\sim\operatorname{Binomial}(r-1,\nu).
 \label{eq:app-runtransform}
\end{align}
For $0<\nu<1$, $M_r\to\infty$ in probability.  Given $\epsilon>0$,
choose $m_0$ such that $s_m-s_\infty<\epsilon$ for $m\ge m_0$.  Then
\begin{equation}
 0\le\mathbb E[s_{M_r}]-s_\infty
 \le\epsilon+\Pr(M_r<m_0),
\end{equation}
and therefore
\begin{equation}
 \lim_{r\to\infty}\lim_{L\to\infty}
 \frac{\overline R_w^{(1)}(r;L)}{\Rsec(L)}=s_\infty.
 \label{eq:app-runcriterion}
\end{equation}
This criterion concerns the single-up-particle channel.  It requires a
nonzero size-biased fraction of up-spin ranks in diverging runs; one
diverging run containing $o(N_\uparrow)$ spins need not produce order.

\subsection{Typical fixed-composition words}
\label{app:typicalstats}

Choose $w$ uniformly among words with $N_\uparrow$ up spins.
For a fixed window of $a=m+1$ ranks, the probability of an all-up window is
$(N_\uparrow)_a/(N)_a$.  Hence
\begin{equation}
 \mathbb E_w A_m=(N-m)\frac{(N_\uparrow)_{m+1}}{(N)_{m+1}}.
 \label{eq:app-runmean}
\end{equation}
Writing $A_m$ as a sum of window indicators gives variance $O(N)$ for
fixed $m$.  Indeed, only $O(Na)$ pairs of windows overlap, each with
bounded covariance.  For disjoint windows the joint probability is
$(N_\uparrow)_{2a}/(N)_{2a}$, which differs from $[(N_\uparrow)_a/(N)_a]^2$ by $O(1/N)$.
The $O(N^2)$ disjoint pairs thus contribute only $O(N)$ in total.
It follows that $A_m/N\to p^{m+1}$ in probability and
$s_m^{(L)}\to p^m$.  A finite union bound at fixed $r$ in
Eq.~\eqref{eq:app-runtransform} yields
\begin{equation}
 \frac{\overline R_w^{(1)}(r;L)}{\Rsec(L)}
 \xrightarrow[L\to\infty]{\mathrm{prob.}}
 (1-\nu+\nu p)^{r-1}.
 \label{eq:app-typicalone}
\end{equation}
The same covariance estimate controls the finite-pattern frequencies in
the source-averaged, many-particle theorem below.

\section{Absence of charged local order in typical fragments}
\label{app:generallocal}

Fix a finite enclosing-interval length $K$, and let
$I_x=\{x,\ldots,x+K-1\}$ and $I_{x+r}$ with integer $r\ge K$
and $L\ge r+K$.
An operator may act on any subset of $I_x$; unused sites carry the identity.
The two intervals are ordered from left to right and leave an untouched
gap of $g=r-K$ sites.  Let $\mathcal O_K$ contain operators supported
in the reference interval $I_1$, translated to $I_x$, with operator norm $\|O\|\le1$
and definite nonzero species charge
$[N_\sigma,O]=q_\sigma O$.
Both $K$ and the source prescription are fixed independently of the word.
In this appendix $L\to\infty$ is taken at fixed $K,r$, and only then is
$r$ increased.
Unless stated otherwise, $\mathbb E_w$ and $\Pr_w$ refer to uniform
fixed-composition words, with $N/L\to\nu\in(0,1)$ and
$N_\uparrow/N\to p\in(0,1)$.

\subsection{A common-input bound for the full operator family}

For a configuration $\alpha$ in a fixed word fragment, let $M(\alpha)$ be the number of particles in the gap and let $j(\alpha)$ be the first gap-particle rank.  For each signed displacement $\delta\in\{-K,\ldots,-1,1,\ldots,K\}$, define the eligibility event
\begin{equation}
 E_\delta(\alpha):\quad
 w_{j(\alpha)+\ell}=w_{j(\alpha)+\ell+\delta},
 \qquad 0\le \ell<M(\alpha),
 \label{eq:app-eligibility}
\end{equation}
with an out-of-range shifted rank counted as failure.  Let $E(\alpha)=\cup_{0<|\delta|\le K}E_\delta(\alpha)$ and let $Q_E$ project onto the eligible input configurations.

Consider first an operator $O$ with definite nonzero species charge $\boldsymbol q=(q_\uparrow,q_\downarrow)$ and set $\delta=q_\uparrow+q_\downarrow$.  If $\delta=0$, cut the chain just after the left interval.  The number of particles to the left of the cut is unchanged by the neutral transfer $O_{x+r}^\dagger O_x$, but the species content of that fixed word prefix changes by $\boldsymbol q$; the distant right operation cannot repair it.  Thus
\begin{equation}
 \Pi_w O_{x+r}^\dagger O_x\Pi_w=0
 \qquad(\delta=0,\ \boldsymbol q\ne\boldsymbol0).
 \label{eq:app-zerodisplacement}
\end{equation}
If $\delta\ne0$, every untouched gap particle changes its rank by $\delta$, so preservation of the final word requires Eq.~\eqref{eq:app-eligibility}.  Hence, with $A_O=\Pi_wO_{x+r}^\dagger O_x\Pi_w$,
\begin{equation}
 A_O=A_OQ_E,
 \qquad
 \operatorname{rank}A_O\le \Tr Q_E.
 \label{eq:app-commoninput}
\end{equation}
Because $\|O\|\le1$, one also has $\|A_O\|\le1$.  All singular values are therefore at most unity, and the projector identities of Appendix~\ref{app:projector} give the operator-uniform bounds
\begin{align}
 F_O(x,x+r;\rho_w)&\le e_w(x,r;K),\nonumber\\
 R_{1,O}(x,x+r;\rho_w)&\le e_w(x,r;K).
 \label{eq:app-rankbound}
\end{align}
where $e_w=\Tr Q_E/d_w$.  The same argument covers an operator with several nonzero charge components and no neutral component: projection to the fixed global charge removes cross terms with unequal endpoint charges, while every surviving process has an input column contained in the same union $E$.

\subsection{Probability of the common eligible set}

In an independent Bernoulli-$p$ word, write $\theta=\max(p,1-p)<1$.  For a fixed nonzero $\delta$, the $M$ equalities in Eq.~\eqref{eq:app-eligibility} form disjoint chains with a total of $M$ edges.  A chain of $n$ vertices is monochromatic with probability $p^n+(1-p)^n\le\theta^{n-1}$, so
\begin{equation}
 \Pr_w(E_\delta\mid M)\le\theta^M.
 \label{eq:app-wordbound}
\end{equation}
For a uniformly sampled fixed-composition word, any fixed finite rank set has the same Bernoulli limit.  The number of particles in the real-space gap $g=r-K$ converges to $M_g\sim\operatorname{Bin}(g,\nu)$, and therefore
\begin{align}
 \limsup_{L\to\infty}\mathbb E_w e_w(x,r;K)
 &\le 2K\,\mathbb E[\theta^{M_g}]\nonumber\\
 &=2K(1-\nu+\nu\theta)^{r-K}
 =2K\lambda^{r-K}.
 \label{eq:app-eligmean}
\end{align}
Endpoint requirements can only reduce the successful set.  The estimate is uniform in a prescribed source position and remains valid after averaging over all allowed sources.  Combining Eq.~\eqref{eq:app-rankbound} with Markov's inequality proves, for every $\epsilon>0$,
\begin{align}
 \limsup_{L\to\infty}\Pr_w\!\left[
 \sup_{O\in\mathcal O_K}\overline F_{O,L}(r;w)>\epsilon\right]
 &\le\frac{2K}{\epsilon}\lambda^{r-K},\nonumber\\
 \limsup_{L\to\infty}\Pr_w\!\left[
 \sup_{O\in\mathcal O_K}\overline R_{1,O,L}(r;w)>\epsilon\right]
 &\le\frac{2K}{\epsilon}\lambda^{r-K}.
 \label{eq:app-probabilitytheorem}
\end{align}
This proves Eq.~\eqref{eq:typical_bound}.  The bound is deliberately restricted to fixed finite $K$; it does not justify a support size that grows with $r$ or $L$, nor does it optimize the source position after inspecting a rare word.

\subsection{One infinite typical word and the order of limits}

For prefixes of a single infinite Bernoulli-$p$ word, an almost-sure
statement follows by retaining the finite-$L$ source average until after
taking the operator supremum.  This changes the probability space.
Set $W=r+K$ and fix an occupation pattern $u\in\{0,1\}^W$,
with particle number $n=\sum_{a=1}^{W}u_a$.
For any fixed local matrix-unit transfer $T_a$ appearing in a configuration-basis expansion of a charged local operator, let $f_{a,u}$ be its successful-transfer
indicator as a function of the $n$ consecutive spin labels in this
window.  A window starting at site $x$ has $x-1$ sites to its left.
If those sites contain $j$ particles, the number of compatible outside
placements is $\binom{x-1}{j}\binom{L-W-x+1}{N-n-j}$.
Summing over all source positions uses
\begin{equation}
 \sum_{x=1}^{L-W+1}\binom{x-1}{j}\binom{L-W-x+1}{N-n-j}
 =\binom{L-W+1}{N-n+1}.
 \label{eq:app-patternvandermonde}
\end{equation}
Thus the contribution of $u$ to $\overline R_{1,T_a,L}$ is exactly
\begin{equation}
 \frac{\binom{L-W}{N-n}}{\binom LN}
 \frac1{N-n+1}\sum_{j=0}^{N-n}
 f_{a,u}(w_{j+1},\ldots,w_{j+n}).
 \label{eq:app-patternaverage}
\end{equation}
Inadmissible patterns are omitted; all patterns become admissible at
large $L$ for fixed $W$ and $0<\nu<1$.
The prefactor tends to $\nu^n(1-\nu)^{W-n}$.  The pointwise ergodic theorem~\cite{Birkhoff1931Ergodic}, applied to the Bernoulli shift, makes
the empirical factor tend almost surely to $\mathbb E f_{a,u}$, bounded
by $\theta^{M(u)}$, where $M(u)$ is the number of gap particles.
Summing over the finitely many $u$ gives $\lambda^{r-K}$.
Ordinary limits also exist uniformly over the full operator family.
For fixed $W$, fixing the exterior configuration decomposes
$\Pi_w X\Pi_w$ into finite blocks $P_vX_W(O)P_v$, where $P_v$ projects
onto all placements in the window of the consecutive spin word $v$.
The neutral product $X=O_{x+r}^\dagger O_x$ acts within this window,
including any fermionic sign string.  Additivity of the squared
Hilbert--Schmidt norm and the trace norm over these blocks expresses
both $\overline R_1$ and $\overline F$ as finite sums with the occupation
prefactors and empirical word frequencies in
Eq.~\eqref{eq:app-patternaverage}.  Their coefficients converge, and the
block norms are bounded uniformly for $\|O\|\le1$.  Thus both correlators
converge uniformly in $O$, so their suprema have ordinary limits.
Applying the common-input bound before the limit gives
\begin{align}
 \lim_{L\to\infty}\sup_{O\in\mathcal O_K}
 \overline R_{1,O,L}(r;w)&\le 2K\lambda^{r-K},\nonumber\\
 \lim_{L\to\infty}\sup_{O\in\mathcal O_K}
 \overline F_{O,L}(r;w)&\le 2K\lambda^{r-K}
 \label{eq:app-almostsure}
\end{align}
on a common probability-one set.  There are countably many finite
$K,r$, finite patterns, and local transfer types, so the same set covers every fixed
finite interval length.  In particular,
almost surely for every finite $K$,
\begin{equation}
 \lim_{r\to\infty}\lim_{L\to\infty}\sup_{O\in\mathcal O_K}
 \overline F_{O,L}(r;w)=0.
 \label{eq:app-almostsurelimit}
\end{equation}
Uniform convergence justifies taking the operator supremum before the
thermodynamic limit.

For exactly fixed-composition words, the finite-pattern frequency in
Eq.~\eqref{eq:app-patternaverage} has variance $O(1/N)$: overlapping
windows contribute $O(N)$ bounded covariances, and disjoint windows have
covariance $O(1/N)$.  The latter follows by expanding their joint
hypergeometric spin probabilities at fixed window size.  The same
finite-pattern argument therefore yields convergence in probability,
uniformly in the operator family.  The correlators and their suprema
are bounded, so their expectations converge as well.
The two probability spaces are distinct.  Neither statement optimizes
over word-dependent rare spatial locations or over supports whose
enclosing diameter grows with the separation.

\section{Periodic words and the six block-transfer curves}
\label{app:periodic}

\subsection{A phase-resolved count for any block motif}

Let $B$ be a primitive block of length $k$ and take the finite words
$w^{(L)}=B^{N(L)/k}$, with $k\mid N(L)$ and $N(L)/L\to\nu$.
Their composition is fixed by $p=n_\uparrow(B)/k$, where
$n_\sigma(B)$ counts species $\sigma$ in $B$.
A transferred motif $C=(\tau_1,\ldots,\tau_\ell)$ has length $\ell$, with
\begin{equation}
 O_i^{(C)}=\widetilde c_{i+\ell-1,\tau_\ell}\cdots
                   \widetilde c_{i,\tau_1},\qquad
 X_{i,r}^{(C)}=O_{i+r}^{(C)\dagger}O_i^{(C)}.
 \label{eq:app-motifoperator}
\end{equation}
The charge is
$\boldsymbol q(C)=(-n_\uparrow(C),-n_\downarrow(C))$ and
$\delta=-\ell$.  The blocks are disjoint for $r\ge\ell$, with
$g=r-\ell$ gap sites.  The periodic source average at fixed $r$ is
\[
 \overline R_C(r;B)=\lim_{L\to\infty}\overline R_C(r;w^{(L)},L).
\]

For phase $a\in\{0,\ldots,k-1\}$, let $V_a(n)$ be the first $n$
spins of the periodic word beginning at phase $a$, and let $W_a(m)$ be
the $m$ spins immediately after its first $\ell$ spins.
Once the source is occupied and the target empty, an applicable transfer
requires $V_a(\ell)=C$.  The compressed word changes from $CW_a(m)$ to
$W_a(m)C$.  Write $\chi_C(a,m)=1$ when both conditions hold, and
zero otherwise.  This gives the exact thermodynamic source average
\begin{align}
 \overline R_C(r;B)={}&\frac{\nu^\ell(1-\nu)^\ell}{k}
 \sum_{a=0}^{k-1}\sum_{m=0}^{g}\binom gm\nonumber\\
 &\times\nu^m(1-\nu)^{g-m}\chi_C(a,m).
 \label{eq:app-phasecount}
\end{align}
This count applies to every positive motif length $\ell$.
The two endpoint occupation factors come from the Bernoulli-$\nu$
limit at fixed $r$.  The factor $1/k$ is the rank-phase frequency,
which also follows directly from the exact source-average formula
\eqref{eq:app-patternaverage}: averages of a periodic local indicator
over consecutive starting ranks converge to the uniform average over
the $k$ phases.  Thus no independence assumption about a fixed source
rank is needed for the source-averaged result.
Each transfer is a basis partial isometry, so $\overline R_C$ also equals
its source-averaged fidelity by Eq.~\eqref{eq:app-successcount}.

For a finite word $w$ with $L\ge r+\ell$, let $j$ count the particles
before the source block.  The indicator $\chi_C(j,m;w)$ is one when
$w_{j+1}\cdots w_{j+\ell}=C$ and this block commutes with the following
$m$-spin subword under concatenation.  Sum over
$0\le j\le N-\ell$ and $0\le m\le r-\ell$:
\begin{align}
 \overline R_C(r;w,L)={}&\frac1{(L-r-\ell+1)\binom LN}\nonumber\\
 &\times\sum_{i=1}^{L-r-\ell+1}\sum_{j,m}
 \chi_C(j,m;w)\binom{i-1}{j}\nonumber\\
 &\times\binom{r-\ell}{m}
 \binom{L-i-r-\ell+1}{N-\ell-j-m}.
 \label{eq:app-finiteblock}
\end{align}
Illegal word windows and binomial coefficients contribute zero.
This formula fixes all boundary and contact conventions before taking
the thermodynamic limit.

\subsection{Primitive-block order and its minimum support}

For $C=B$ and $\ell=k$, only one source phase matches.  The remaining
word equality is satisfied precisely when $m=0\pmod{k}$.
For necessity, the elementary commuting-word identity $BW=WB$ implies
that $B$ and $W$ are powers of one common word~\cite{Lothaire1997Words}.  This follows by
repeated prefix cancellation: the shorter word is a prefix of the
longer, and deleting that prefix reduces the sum of the two lengths
while preserving commutation.  Since $B$ is primitive, the common
word is $B$, and $|W|$ is a multiple of $k$.  Sufficiency follows by
moving one complete period through a sequence of complete periods.
For $r\ge k$, set $g=r-k$ and
$M_g\sim\operatorname{Binomial}(g,\nu)$.  A roots-of-unity projection gives
\begin{align}
 \overline R_B(r;B)
 &=\frac{\nu^k(1-\nu)^k}{k}
       \Pr[M_g=0\pmod{k}]\nonumber\\
 &=\frac{\nu^k(1-\nu)^k}{k^2}
   \sum_{h=0}^{k-1}
       (1-\nu+\nu e^{2\pi i h/k})^{r-k}.
\label{eq:app-primitivecurve}
\end{align}
At contact $g=0$, $M_0=0$ deterministically, so each zeroth power
in the sum is one, including a zero base.  Thus
$\overline R_B(k;B)=\nu^k(1-\nu)^k/k$.
For $0<\nu<1$, every term with $h\ne0$ has modulus less than one
before exponentiation.  Hence
\begin{equation}
 \lim_{r\to\infty}\overline R_B(r;B)
 =\frac{\nu^k(1-\nu)^k}{k^2}>0.
 \label{eq:app-primitiveplateau}
\end{equation}

To establish minimality, consider a definite-charge operator in a fixed
enclosing interval of length $K$.  A nonzero displacement $\delta$
that is not a period of the infinite word violates
$w_j=w_{j+\delta}$ somewhere in every $k$ consecutive ranks.
Equation~\eqref{eq:app-eligibility} then permits a successful transfer
only if the gap contains fewer than $k$ particles.  Here $g=r-K$ and
$M_g\sim\operatorname{Binomial}(g,\nu)$, so the probability is bounded by
\begin{equation}
 \Pr[M_g<k]=\sum_{m=0}^{k-1}\binom gm\nu^m(1-\nu)^{g-m}
 \xrightarrow[g\to\infty]{}0.
 \label{eq:app-periodminimal}
\end{equation}
The common-input bound excludes order in that channel.  Every period of
a bi-infinite primitive period-$k$ word is a multiple of $k$, so a
nonzero ordered displacement must satisfy
$\delta\in k\mathbb Z\setminus\{0\}$.
The case $\delta=0$ with nonzero species charge is excluded by
Eq.~\eqref{eq:app-zerodisplacement}.
With $s=|\operatorname{supp}O|$, the particle number changes by at most
$s$, so $K\ge s\ge|\delta|\ge k$.  For a sum of nonzero charge
components, order requires at least one component with such a $\delta$;
the same support bound follows.  The block operator saturates the bound
for this periodic family.
Other word families remain to be classified, and other operators may
have the same minimum support.

\subsection{Species-charge selection inside a periodic word}

Let the primitive block $B$ contain
$\boldsymbol b=(n_\uparrow(B),n_\downarrow(B))$.  Cut just after the left operator interval.  If $n_-$ particles lie to the left initially, a word-preserving operation with local particle-number change $\delta$ and species charge $\boldsymbol q$ must satisfy
\begin{equation}
 \boldsymbol q=\boldsymbol C_w(n_-+\delta)-\boldsymbol C_w(n_-),
 \label{eq:app-prefixcharge}
\end{equation}
where $\boldsymbol C_w(n)$ counts the two species in the first $n$ ranks.  Long-range order in a primitive period-$k$ background requires $\delta=mk$ with $m\ne0$.  Every $k$ consecutive ranks contain the same species vector $\boldsymbol b$, and hence
\begin{equation}
 \boldsymbol q=m\boldsymbol b,
 \qquad m\in\mathbb Z\setminus\{0\}
 \label{eq:app-chargelattice}
\end{equation}
is a necessary charge-selection rule.  It is not sufficient by itself because endpoint occupations and matrix elements must still match.

A useful contrast is the local spin-transfer operator $O_i=S_i^+$, with $\boldsymbol q=(1,-1)$ and $\delta=0$.  The prefix argument gives
\begin{equation}
 \Pi_wS_j^-S_i^+\Pi_w=0,
 \qquad F_{S^+}=R_{1,S^+}=0
 \label{eq:app-spinfragment}
\end{equation}
for every fixed word and separated sites.  In the complete fixed-charge projector the transfer succeeds whenever the source is down and the target is up, so
\begin{align}
 F_{S^+}(i,j;\rho_{\rm sec})
 &=R_{1,S^+}(i,j;\rho_{\rm sec})\nonumber\\
 &=\frac{N_\uparrow N_\downarrow}{L(L-1)}
 \longrightarrow\nu^2p(1-p).
 \label{eq:app-spinsector}
\end{align}
At half filling and spin balance the limit is $1/16$.

\subsection{Alternating word: one, two, and four particles}

Here $B=UD$, and we suppress $B$ in $\overline R_C(r;B)$.
The up-run length is one.  Only $m=0$ contributes
to the single-particle count, so
\begin{equation}
 \overline R_U(r)=\frac{\nu(1-\nu)}2(1-\nu)^{r-1}.
 \label{eq:app-udsingle}
\end{equation}
For $C=UD$ and $C=UDUD$, with $\ell=2$ and $4$, respectively, one of
the two source phases matches and word preservation requires an even
number of gap particles.  Both curves follow from
\begin{equation}
 \overline R_C(r)=\frac{\nu^\ell(1-\nu)^\ell}{4}
             [1+(1-2\nu)^{r-\ell}],\qquad r\ge \ell.
 \label{eq:app-udeven}
\end{equation}
At half filling,
\begin{align}
 \overline R_U(r)&=\frac{2^{-(r-1)}}8,\qquad r\ge1,\nonumber\\
 \overline R_{UD}(r)&=\frac1{64},\qquad r>2,\nonumber\\
 \overline R_{UDUD}(r)&=\frac1{1024},\qquad r>4.
 \label{eq:app-udcurves}
\end{align}
The contact values are $\overline R_{UD}(2)=1/32$ and
$\overline R_{UDUD}(4)=1/512$, because an empty gap satisfies the parity condition
with probability one.  In the four-particle plateau, one factor $1/2$
comes from the source phase and another from the asymptotic even-gap
probability.  The primitive word period is still two.

\subsection{Period-four word: the lower channels and the oscillating plateau}

For $B=UUDD$, we also suppress the background label.
For $C=U$, two of four phases allow a zero-particle
gap.  Only the first up-spin phase allows one gap particle, and no phase
allows two or more.  With $g=r-1$,
\begin{equation}
 \overline R_U(r)=\nu(1-\nu)\left[
 \frac12(1-\nu)^g+\frac14 g\nu(1-\nu)^{g-1}\right].
 \label{eq:app-uuddsingle}
\end{equation}
The second term is absent at $g=0$.
For $C=UD$, only one phase matches. The subsequent period-four sequence cannot commute with this motif for any positive gap occupation, so only $m=0$ survives:
\begin{equation}
 \overline R_{UD}(r)=\frac{\nu^2(1-\nu)^2}{4}(1-\nu)^{r-2}.
 \label{eq:app-uuddpair}
\end{equation}
At half filling these become
\begin{equation}
 \overline R_U(r)=\frac{r+1}{2^{r+3}},\qquad
 \overline R_{UD}(r)=\frac1{2^{r+4}}.
 \label{eq:app-uuddshort}
\end{equation}
The motifs $UU$, $DD$, and $DU$ each occur at one phase and also fail
the word equality for every positive gap occupation.  They therefore
have the same two-particle curve.  In particular, choosing $UD$ in
both periodic words compares the same species-charge channel and
still gives a plateau for the alternating word and decay for $UUDD$.

For $C=UUDD$, Eq.~\eqref{eq:app-primitivecurve} applies with $k=4$.
At $\nu=1/2$, the four roots contribute $1$, $0$, and
$2^{-1/2}e^{\pm i\pi/4}$.  For $g=r-4>0$ this gives
\begin{equation}
 \overline R_{UUDD}(r)=\frac{1+2\,2^{-g/2}\cos(\pi g/4)}{4096},
 \qquad r>4.
 \label{eq:app-uuddquartet}
\end{equation}
The separate contact value is $\overline R_{UUDD}(4)=1/1024$.
The exact thermodynamic count produces the damped oscillation about
$1/4096$ even in the absence of finite-size or sampling effects.

For comparison, the complete-sector projector gives, for a prescribed
$UD$ source and an empty two-site target,
\begin{equation}
 R_{\mathrm{sec},UD}(L)
 =\frac{N_\uparrow N_\downarrow(L-N)(L-N-1)}
        {L(L-1)(L-2)(L-3)}.
 \label{eq:app-sectorpair}
\end{equation}
This follows by fixing the four endpoint occupations and counting the
remaining multinomial configurations.  Its half-filled, balanced limit
is $1/64$, equal to the pair plateau in the alternating fragment.

\begin{figure*}[!t]
\centering
\includegraphics[width=\textwidth]{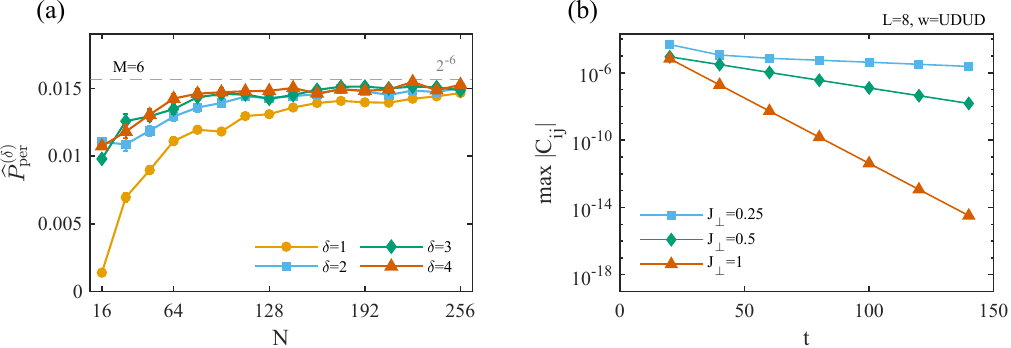}
\caption{Word-statistics and ordinary-coherence controls. (a) Balanced words ($p=1/2$) with $M=6$, $\delta=1,2,3,4$, and $N=16,32,\ldots,256$. The mean fraction of valid periodic rank windows approaches $2^{-6}$ as $N$ grows. The $N=16$ means are exact averages over all $\binom{16}{8}=12870$ balanced words and have no sampling error bars; at each larger $N$, the means are estimated from $5000$ sampled words and shown with standard-error bars. Markers give the values at each listed $N$, and solid lines connect adjacent values as guides to the eye. This limit removes fixed-composition corrections and does not measure decay of a spatial correlator; Eq.~\eqref{eq:app-finiteNperiodicity} gives the exact finite-$N$ reference. (b) Maximum ordinary coherence $\max_{|i-j|=4}|C_{ij}(t)|$, with $C_{ij}(t)=\Tr[\rho(t)\widetilde c_{i\uparrow}^\dagger\widetilde c_{j\uparrow}]$, starting from $\rho_{UDUD}$ with the parameters of Fig.~\ref{fig:dynamics}. Values are shown at $t=20,40,\ldots,140$ for $J_\perp=0.25,0.5,1$. The last plotted point is $t=140$, and the horizontal axis ends at $t=150$. Solid lines connect adjacent plotted time points as guides to the eye. The initial values and the $J_\perp=0$ series are exactly zero and omitted from the log axis. The smallest values are near floating-point precision and are not used to fit a decay rate.}
\label{fig:controls}
\end{figure*}

\section{Numerical evaluation}
\label{app:numerics}

\subsection{Density-matrix evolution at \texorpdfstring{$L=8$}{L=8}}

The plotted calculation uses $L=8$, $t_{\mathrm h}=1$, $J_z=0.7$,
$\gamma=1$, and $N_\uparrow=N_\downarrow=2$.
The transverse term is normalized as
$J_\perp\sum_i(S_i^+S_{i+1}^-+S_i^-S_{i+1}^+)$.
The complete-sector basis contains
$\binom84\binom42=420$ configurations and each word fragment has
$\binom84=70$.  The Hamiltonian and transfer matrices are assembled in
this fixed-charge basis.  With column-wise vectorization, the generator
has Hamiltonian part $-i(\mathbf1\otimes H-H^{\mathsf T}\otimes\mathbf1)$.
Write $\mathcal D_{\rm dep}=\gamma\sum_{i,\sigma}\mathcal D[n_{i\sigma}]$
for total species dephasing and
$n_{i\sigma}^{\alpha}=\langle\alpha|n_{i\sigma}|\alpha\rangle\in\{0,1\}$.
Its matrix elements obey
\begin{equation}
 [\mathcal D_{\rm dep}(\rho)]_{\alpha\beta}
 =-\frac\gamma2\sum_{i,\sigma}
       (n_{i\sigma}^{\alpha}-n_{i\sigma}^{\beta})^2
       \rho_{\alpha\beta}.
 \label{eq:app-dephasingmatrix}
\end{equation}
The vectorized density matrix is propagated with the sparse
matrix-exponential action \texttt{expm\_multiply}~\cite{AlMohyHigham2011}.
This avoids trajectory sampling for the $L=8$ evolution.
At each sampled time, we evaluate the source-averaged transient signal
from the full trace ratio
\begin{equation}
 \overline R_2^{(1)}(r,t;L)=\frac1{L-r}\sum_{i=1}^{L-r}
 \frac{\Tr[X_{i,r}^{(1)}\rho(t)X_{i,r}^{(1)\dagger}\rho(t)]}
      {\Tr[\rho(t)^2]}.
 \label{eq:app-transientr2}
\end{equation}
At $r=4$, the calculation averages over the eight directed
pairs $|i-j|=4$.  This equals the four-source expression above because
cyclicity gives $\Tr(\rho X\rho X^\dagger)
=\Tr(\rho X^\dagger\rho X)$.
The linear diagnostic is the maximum of
$|\Tr(\rho\widetilde c_{i\uparrow}^\dagger
\widetilde c_{j\uparrow})|$ over those same eight directed pairs.
The signal $g_2(t)$ divides this source average at $r=L/2$ by
$\Rsec(L)=1/7$ at this size.
For the alternating fragment at $L=8$, its exact initial value is
$g_2(0)=1/\binom63=1/20$; the connected-sector endpoint is $1$.
The time-dependent panels in Figs.~\ref{fig:dynamics}(a) and \ref{fig:controls}(b) start from $\rho_0=\Pi_{UDUD}/70$.
Figure~\ref{fig:controls}(b) shows results at the seven times $t=20,40,\ldots,140$ for
$J_\perp=0.25,0.5,1$, with the horizontal axis extending to $t=150$. Figure~\ref{fig:dynamics}(a) shows
$t=0,1,\ldots,120$ for $J_\perp=0,0.25,0.5,1$, with $121$ evaluations per curve.
The late-time comparison starts from each of the six balanced word
projectors and uses $J_\perp=1$ at $t=120$.
In both figures, lines connect data at adjacent times only as guides to the eye; no additional
interpolated, smoothed, or fitted points are used. Very small linear-correlator tails are shown
without a numerical floor and are not used to fit decay exponents.

A cross-check between sparse exponential propagation and a variable-step
backward-differentiation solver gave agreement to
$5.39\times10^{-12}$ in the endpoint $\overline R_2^{(1)}$ for the species bath
at $J_\perp=1$.  The corresponding density-matrix Frobenius difference
was $1.66\times10^{-8}$; recorded trace and Hermiticity errors were at
most $7.22\times10^{-15}$ and $2.18\times10^{-17}$, respectively.
These solver checks assess numerical agreement; the deterministic
curves have no sampling error bars.

\subsection{Periodicity statistics and exact block curves}

For a word of length $N$ and positive $\delta$, there are
$n_{\rm win}=N-M-\delta+1$ valid rank windows.  When $n_{\rm win}\ge1$,
their periodicity density is
\begin{equation}
 P_{\mathrm{per}}^{(\delta)}(M;w)
 =\frac1{n_{\rm win}}\sum_{j=1}^{n_{\rm win}}
 \prod_{\ell=0}^{M-1}
 \mathbf1[w_{j+\ell}=w_{j+\ell+\delta}].
 \label{eq:app-perioddensity}
\end{equation}
The plotted word mean is
\[
 \widehat P_{\mathrm{per}}^{(\delta)}(M;N)
 =\frac1{n_{\rm samp}}\sum_{b=1}^{n_{\rm samp}}
 P_{\mathrm{per}}^{(\delta)}(M;w^{(b)}).
\]
It is a word statistic, distinct from the spatial correlator $R_1$.
Negative displacements reduce to the same step $|\delta|$ by relabeling
the rank indices.  For independent balanced spins the expectation
is exactly $2^{-M}$ by the equality-chain count.  Exact composition
produces finite-$N$ corrections.  These corrections can be computed without
sampling.  Let the word contain exactly $U$ up spins, set $d=|\delta|$ and
$W=M+d\le N$, and let the $d$ equality chains have lengths
\begin{equation}
 n_a=1+\left\lfloor\frac{W-a}{d}\right\rfloor,
 \qquad a=1,\ldots,d.
 \label{eq:app-chainlengths}
\end{equation}
Each chain must be entirely up or entirely down.  Summing over the subset
$S$ of chains chosen to be up gives the exact fixed-composition mean
\begin{equation}
 \mathbb E_w P_{\mathrm{per}}^{(d)}(M;w)
 =\frac{\displaystyle\sum_{S\subseteq\{1,\ldots,d\}}
 \binom{N-W}{U-\sum_{a\in S}n_a}}
 {\binom NU}.
 \label{eq:app-finiteNperiodicity}
\end{equation}
Out-of-range binomial coefficients are zero.  For the plotted
$d=1,2,3,4$ the sum contains at most sixteen terms and tends to
$2^{-M}$ at fixed $M,d$ as $N\to\infty$.  Thus the horizontal $2^{-6}$
reference in the size scan is the large-$N$ value, while
Eq.~\eqref{eq:app-finiteNperiodicity} gives the exact finite-$N$ benchmark.
Each word is one statistical unit: overlapping windows are averaged
within that word before computing the mean and standard deviation
across words. The size scan uses all $\binom{16}{8}=12870$ balanced words at $N=16$
and 5000 balanced words at each $N=32,48,\ldots,256$. The sampled words are generated
with MT19937 seed $220000+N$.
The bars at sampled sizes are $s/\sqrt{n_{\rm samp}}$, where $s$ is the sample standard deviation across words; they estimate Monte Carlo uncertainty. At $N=16$, the means are exact and have no sampling error bars. Figure~\ref{fig:controls}(a) gives the fixed-$M$ size scan. Figure~\ref{fig:typical}(a) is the thermodynamic analytic curve from Eq.~\eqref{eq:onebody_typical}, without finite-size counting or trajectory data.

The projector results in Figs.~\ref{fig:typical}(a) and
\ref{fig:periodic} use $J_\perp=0$ and the thermodynamic limit at
$\nu=p=1/2$.  These static counts are independent of $J_z$ and of
the positive dephasing rate $\gamma$.  Figure~\ref{fig:typical}(a)
draws the analytic formula over $1\le r\le20$; only integer $r$ is a
physical separation.  The word statistics in
Figs.~\ref{fig:typical}(b) and \ref{fig:controls}(a) depend only on
$N,p,M,\delta$ and the word ensemble; they do not require a spatial
filling, a chain length $L$, or dynamical parameters.

The six periodic block curves are evaluated from
Eq.~\eqref{eq:app-phasecount} by integer binomial sums and exact rational
arithmetic at $\nu=1/2$. The plotted source data contain every integer
separation $r=\max(3,\ell+1),\ldots,20$. The four-site curves begin at
$r=5$; their $r=4$ contacts are reported explicitly above. The solid lines
connect neighboring integer separations only as guides to the eye; no
smoothed, fitted, or extrapolated values are used.
Independent enumeration of 268680 phase/occupation windows reproduced
the exact formula in 52 cases. Further checks covered the phase-and-gap formulas
in 1128 cases at $\nu=1/3,1/2,2/3$ and 252 two-spin-motif comparisons for
the period-four word.  These checks concern the six explicit curves.

\bibliography{references}
\end{document}